\documentclass[aps,prb,reprint,groupedaddress,showpacs]{revtex4-1}
\usepackage{graphicx}
\usepackage{dcolumn}
\usepackage{bm}
\newcommand{\lsim}{\raisebox{-0.7ex}{$\;\stackrel{\textstyle <}{\textstyle\sim}\;$}}
\begin{document}
\title{Moving zero-gap Wannier-Mott excitons in graphene}

\author{M.M. Mahmoodian}
\email{mahmood@isp.nsc.ru}
\affiliation{Institute of Semiconductor Physics, Siberian Branch, Russian Academy of Sciences, Novosibirsk, 630090 Russia}
\altaffiliation[Also at ]{Novosibirsk State University, Novosibirsk, 630090 Russia}

\author{M.V. Entin}
\email{entin@isp.nsc.ru}
\affiliation{Institute of Semiconductor Physics, Siberian Branch, Russian Academy of Sciences, Novosibirsk, 630090 Russia}


\begin{abstract}
We demonstrate the possibility of existence of indirect moving Wannier-Mott excitons in graphene. Electron-hole binding is conditioned by the trigonal warping of conic energy spectrum. The binding energies are found for the lowest exciton states. These energies essentially depend on the value and direction of exciton momentum and vanish when the exciton momentum tends to the conic points. The ways to observe the exciton states are discussed. The opportunity of experimental observation of zero-gap excitons by means of external electron scattering is examined.
\end{abstract}


\maketitle

\section{Introduction}

Graphene is a remarkable material due to its unusual conic electron energy spectrum. One of the important consequences of the spectrum singularity is zero effective mass of electrons near the conic point that results in a high mobility of charge carriers in graphene. Just this fact makes graphene very perspective for electronics. Another important property of freely suspended graphene is a strong e-e interaction that essentially distorts the single-particle spectra.

The purpose of the present paper is an envelope-approximation study of possibility of the Wannier-Mott excitons formation near the conic point in a neutral graphene. There were a lot of studies considered the excitons in graphene. In all these papers the term ''exciton'' is used for many-body (''excitonic'') effects \cite{yang1,yang2}, exciton insulator with full spectrum reconstruction, or exciton-like singularities originating from saddle points (van Hove singularity) of the single-particle spectrum \cite{klit} where the real electron-hole binding does not occur. On the contrary, our goal is pair bound states of electrons and holes.

It is a widely accepted opinion that zero gap in graphene forbids the Mott exciton states (see, i.g., \cite{ratnikov}). This statement is valid in the conic approximation only. Taking the real spectrum structure into account makes the electron-hole binding possible. Namely, it will be shown that the exciton is conditioned by the trigonal corrections to the conic spectrum.

The trigonal warping of the energy spectrum is responsible for the electron-electron scattering selection rules in the neutrality point and, consequently, a specific mechanism of the valley currents in graphene \cite{golub}. Here we shall demonstrate that the presence of the warping drastically affects the possibility of electron-hole binding also.

Graphene has no energy gap, hence we consider an electron-hole bound state with the energy lying in the free electron-hole continuum. Such exciton states exist in usual semiconductors with gaps. However, there is an important difference between quadratic energy spectrum of particles $p^2/2m$ and the conic spectrum in graphene $v p$, where $v$ is electron velocity in the conic approximation. Speaking in classical language, for quadratic energy spectrum the attraction leads to oscillations of particles captured in a mutual well. On the contrary, for the conic spectrum the character of motion of two particles is different, namely, the distance between particles with parallel momenta remains constant due to equality of velocities. The correct description of motion needs going beyond the conic approximation. In this case the series expansion of the pair kinetic energy in the relative momentum determines the finite inverse mass (second derivative of energy with respect to the relative momentum) that leads to the variation of the e-h distance. The pair remains bound if the sign of this mass is positive. This picture remains valid in the quantum consideration. Owing to the dependence of the sign of the mass on the exciton momentum angle, the binding exists in the particular sectors of the total momentum space and is forbidden in others.

We consider graphene with weakly interacting charged carriers, when the interaction constant $g=e^2/v\chi\ll 1$, where $e$ is electron charge, $\chi$ is the half-sum of dielectric constants of surrounding media, $\hbar=1$. This condition guarantee the smallness of the binding energy as compared with the kinetic energy (in other words, applicability of envelope function approximation) and negligibility of many-body effects.

\section{Exciton states}
We consider graphene with primitive translation vectors ${\bf a}=a(1,0)$ and ${\bf b}=a(-1/2,\sqrt{3}/2)$, where $a=0.246$ nm is the lattice constant. The starting point is the tight-binding electron Hamiltonian
\begin{equation}\label{He}
H_e({\bf p})=\left(
      \begin{array}{cc}
        0 & \Omega_{\bf p} \\
        \Omega^*_{\bf p} & 0 \\
      \end{array}
    \right),~~~~~~\Omega_{\bf p} =-\gamma\sum_ie^{-i{\bf pl}_i}.
\end{equation}
Here $\gamma=2v/a\sqrt{3}$, $l_1=(0,1)a/\sqrt{3}$, $l_2=(-3,-\sqrt{3})a/6$, $l_3=(3,-\sqrt{3})a/6$. The spectrum reads as $\varepsilon_\pm({\bf p})=\pm\gamma\sqrt{1+4\cos\frac{ap_x}{2}\cos\frac{\sqrt{3}ap_y}{2}+4\cos^2\frac{ap_x}{2}}.$ The primitive reciprocal lattice has basis vectors ${\bf a}^*=(2\pi/a)(1,1/\sqrt{3}),~~~ {\bf b}^*=(2\pi/a)(0,2/\sqrt{3})$.

In the considered case the exciton states are composed from a narrow group in the momentum space near the points $K$ and $K'$, where ${\bf p}=\pm {\bf K}$, ${\bf K}=(1,0)4\pi/3a$ and the single-electron spectrum is conic: $\varepsilon_\pm ({\bf p})=\pm v|{\bf p- K}|$ and $\varepsilon_\pm ({\bf p})=\pm v|{\bf p+K}|$, correspondingly. In the envelope approximation the two-body Hamiltonian is
\begin{equation}\label{Hexc}
H_{ex}=H_e({\bf p}_e)\otimes I_h-I_e\otimes H_h({\bf p}_h)+I_e\otimes I_hV({\bf r}_e-{\bf r}_h),
\end{equation}
where $H_h({\bf p})=-H_e({-\bf p})$ is the hole Hamiltonian, ${\bf r}_{e,h}$ and ${\bf p}_{e,h}$ are the coordinates and momenta of electron and hole, $V({\bf r})=-e^2/\chi r$ is the potential of the electron-hole interaction. The electron and hole Hamiltonians relate to the electron and the hole subspaces of quantum numbers. Hamiltonian (\ref{Hexc}) has a 4-fold set of eigenstates, one of which belongs to the ordinary exciton, namely, a pair with ''an electron in the conduction band and a hole in the valence band''.

Let us carry out an unitary transform $U=U_e({\bf p}_e)U_h({\bf p}_h)$ of the Hamiltonian (\ref{Hexc}), where $U_e({\bf p}_e)$ and $U_h({\bf p}_h)$ diagonalize $H_e({\bf p}_e)$ and $H_h({\bf p}_h)$, respectively. The smallness of the binding energy means a large size of exciton wave function and the smallness of the spatial derivatives; in that case one can neglect the non-commutativity of $U$ and $V({\bf r}_e-{\bf r}_h)$. As a result, we get to the Hamiltonian:
\begin{equation}\label{HH}
H_{ex}=\varepsilon_+({\bf p}_e)+\varepsilon_+({\bf p}_h)+V({\bf r}_e-{\bf r}_h).
\end{equation}

\begin{figure}
\includegraphics[width=4.5cm]{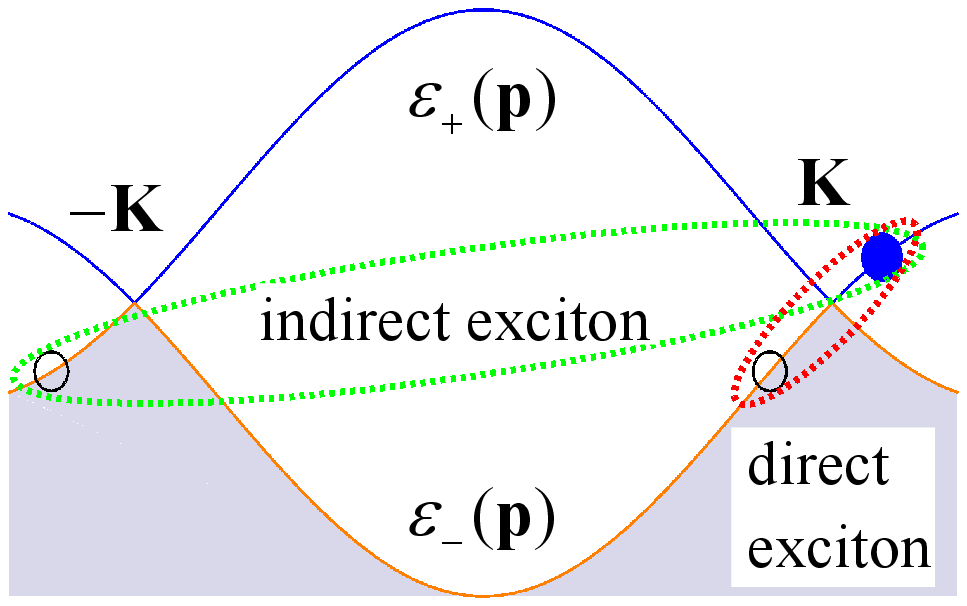}
\includegraphics[width=6cm]{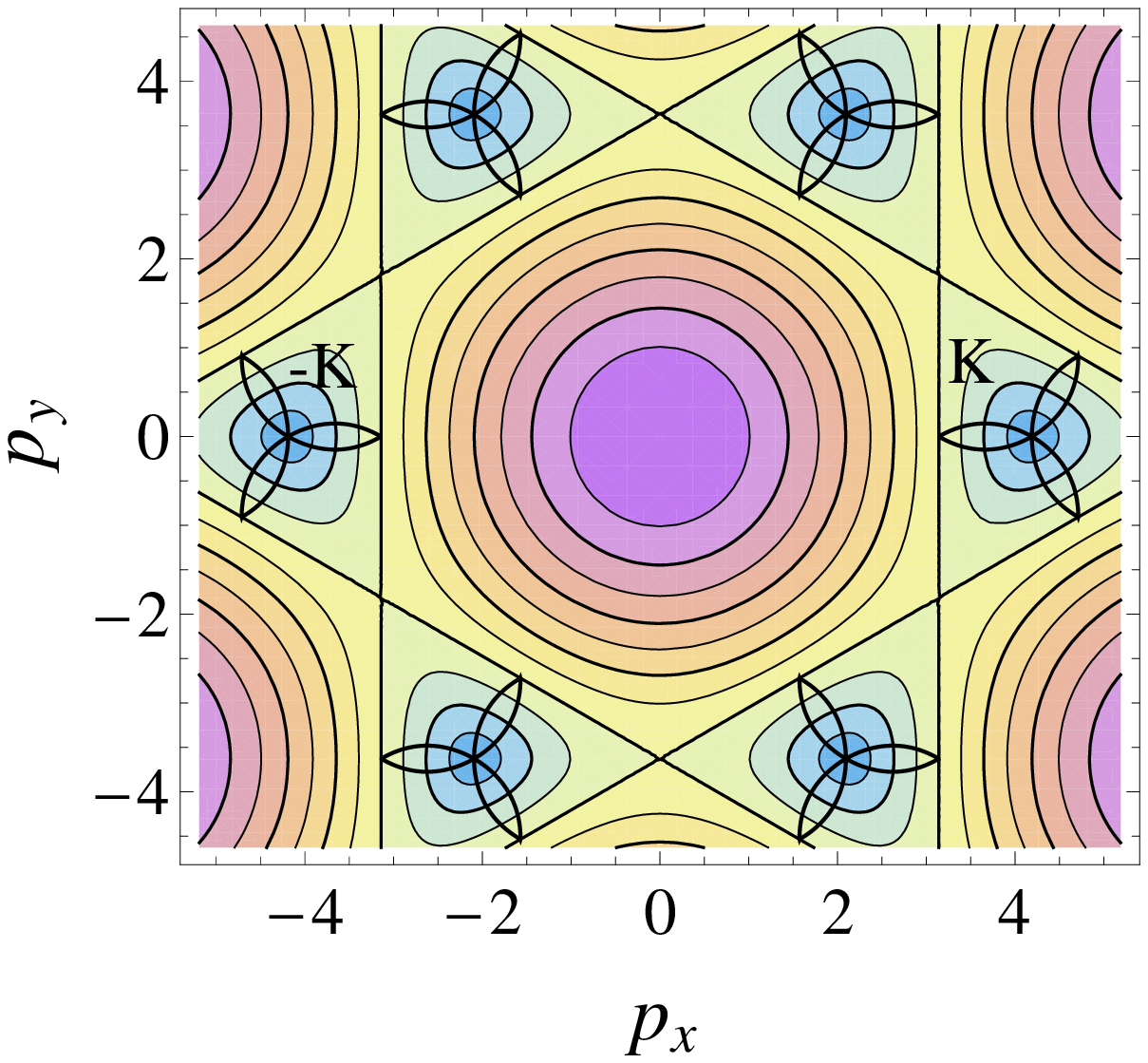}
\caption{(Color online) Top: direct and indirect exciton formation in terms of single-particle spectrum. Bottom: relief of the single-electron spectrum. Energy runs from 0 in the conic points to 3$\gamma$ in the center. The trefoils bound domains of electron momentum ${\bf p}_e$  for which indirect excitons exist, in accordance with Eq.~(\ref{exist}).}
\label{fig1}
\end{figure}

The quantities ${\bf p}_e$ and ${\bf p}_h$ can be expressed via pair ${\bf q}={\bf p}_e+{\bf p}_h$ and relative ${\bf p}={\bf p}_e-{\bf p}_h$ momenta. The momenta can be situated near the same ($q\to k\ll K$) (the direct exciton) or near the opposite conic points (${\bf q}=2{\bf K}+{\bf k},$ $k\ll K$) (the indirect exciton) (see fig.~\ref{fig1}).

Expanding the Hamiltonian with respect to $p/k\ll1$, we find for the indirect excitons
\begin{equation}\label{Hexc2}
H_{ex}=2\varepsilon_+({\bf q}/2)+\frac{p_ip_j}{2}\left(\frac1m\right)_{i,j}-\frac{e^2}{\chi r},
\end{equation}
where ${\bf r}={\bf r}_e-{\bf r}_h$. The components of the tensor of inverse masses $(m^{-1})_{i,j}=\nabla_i\nabla_j \varepsilon_+({\bf q}/2)/2$ read
\begin{eqnarray}\label{mass}
(m^{-1})_{x,x}&=&\varepsilon_+^{-3}\big(2-8c_x^4-c_x c_y-12c_x^3 c_y-\nonumber\\
&&2c_y^2-2c_x^2\big(2+c_y^2\big)\big),\nonumber\\
(m^{-1})_{x,y}&=&\varepsilon_+^{-3}\sqrt{3}\left(1+2c_x c_y\right)s_x s_y\\
(m^{-1})_{y,y}&=&-3\varepsilon_+^{-3}c_x\left(2c_x+c_y\right)
\left(1+2c_x c_y\right)\nonumber
\end{eqnarray}

Here $c_x=\cos(q_xa/4), ~~c_y=\cos(\sqrt{3}q_ya/4), ~~s_x=\sin(q_xa/4), ~~s_y=\sin(\sqrt{3}q_ya/4).$
In the conic approximation $k\ll K$, and the Hamiltonian (\ref{Hexc2}) can be written as
\begin{equation}\label{Hexc3}
H_{ex}=E_{kin}+\frac{p_1^2}{2M}+\frac{p_2^2}{2m}-\frac{e^2}{\chi\sqrt{x_1^2+x_2^2}},
\end{equation}
where $E_{kin}=vk-(\mu k^2/2)\cos3\phi_{\bf k}$ is the kinetic energy of a free pair, the coordinate system with basis vectors ${\bf e}_1\equiv{\bf k}/k$ and ${\bf e}_2\bot{\bf e}_1$ is chosen, $\mu=va/(4\sqrt{3})$ is a warping parameter and we have $m=k/v$, $M=\infty$. To obtain finite $M$, one should go beyond the conic approximation. For the indirect exciton with the momenta close to $\nu2{\bf K}$, ($\nu=\pm1$), we have $1/M=\nu\mu\cos3\phi_{\bf k}$, where $\phi_{\bf k}$ is an angle between ${\bf k}$ and ${\bf K}$. The effective mass $M$ is directly determined by the trigonal contributions to the spectrum and is parametrically large: $\eta=M/m=v/(\nu\mu k\cos3\phi_{\bf k})\gg 1$. The sign of $M$ is determined by $\nu\cos3\phi_{\bf k}$. If $\nu\cos3\phi_{\bf k}>0$, the electron and the hole tend to bind, otherwise to run away from each other. Thus, the binding is possible for $\nu\cos3\phi_{\bf k}>0$.

The similar reasoning for the pair from the same valley shows that the trigonal contributions to $1/M$ cancel each other. The contributions of a higher order give, in this case, a negative mass $1/M$: $1/M=-kva^2(7-\cos6\phi_{\bf k})/32$. As a result, the electron-hole binding is forbidden for particles from the same valley.

Let us consider indirect excitons apart from the conic points. The condition of their existence is positively definiteness of the kinetic energy $p_ip_j(1/m)_{i,j}/2$. In other words, both eigenvalues of the tensor $(1/m)_{i,j}$ should be positive. We find from Eq.~(\ref{mass}):
\begin{eqnarray}\label{exist}
&&(2c_x^2+c_xc_y+s_xs_y)<0\wedge(2c_x^2+c_xc_y-s_xs_y)<0\vee\nonumber\\
&&(2c_x^2+c_xc_y+s_xs_y)<0\wedge(1+2c_xc_y)<0\vee\\
&&(2c_x^2+c_xc_y-s_xs_y)<0\wedge(1+c_xc_y)<0\nonumber.
\end{eqnarray}

Fig.~\ref{fig1} shows the domain of exciton existence in the ${\bf p}_e$-space. This domain covers a small part of the Brillouin zone. Here and below the momentum in the Figs. is measured in units of $1/a$.

The problem of Coulomb states for the Hamiltonian (\ref{Hexc2}) can be considered using the strong anisotropy of the energy spectrum. The kinetic energy of the free pair $E_{kin}$ is the main part of the total exciton energy $E_{nN}=E_{kin}-\varepsilon_{nN}$, where $\varepsilon_{nN}>0$ is the binding energy, quantum numbers $n$ and $N$ numerate the exciton state. A large ratio of masses $\eta\gg 1$ reminds of the problem of molecular states, where $m$ and $M$ represent electron and ion masses, correspondingly. So the way of solution is similar to the problem of the molecular levels. First, we cancel $p_1^2/2M$, fix the ''ion'' coordinate ''$x_1$'' and determine the energy terms $-\zeta_{n}(x_1)$, then use these terms as ions potential energy. The total wave function $\Psi_{nN}(x_1,x_2)$ factorizes to the product of "electron" and "ion" wave functions: $\Psi_{nN}(x_1,x_2)=\psi_n(x_2;x_1)\Psi_{nN}(x_1)$. The binding energy $\varepsilon_{nN}$ in the N-th ''ion'' state on the n-th ''electron'' term $\zeta_{n}(x_1)$ is determined by the Schr\"{o}dinger equation for ''ions''
\begin{equation}\label{Hn}
\left(p_1^2/2M-\zeta_{n}(x_1)+\varepsilon_{nN}\right)\Psi_{nN}(x_1)=0.
\end{equation}
The order of magnitude of the lowest levels is the Bohr energy $\epsilon_B=me^4/\chi^2$.

Let us replace the potential of electron-hole interaction by $V({\bf r})\simeq-e^2/\chi(|x_1|+|x_2|)$ which has the same asymptotics. This approach permits to approximate the energy terms at small $x_2$ ($x_2\ll x_1$). The replacement converts the Schr\"{o}dinger equation to an analytically solvable form:
\begin{eqnarray}\label{hy}\nonumber
\frac{1}{2m}\frac{\partial^2\psi_n(x_2;x_1)}{\partial x_2^2}+\frac{e^2}{\chi(|x_2|+|x_1|)}\psi_n(x_2;x_1)=\\\zeta_n(x_1)\psi_n(x_2;x_1).
\end{eqnarray}
The solution of Eq.~(\ref{hy}) satisfying a zero boundary condition at $|x_2|\to\infty$ ($\zeta_n(x_1)>0$) is
\begin{eqnarray}\label{psiy}\nonumber
\psi_n(x_2;x_1)=C_\pm e^{-P_n(x_1)(|x_2|+|x_1|)}(|x_2|+|x_1|)\times\\ U\left[1-(P_n(x_1)a_B)^{-1}, 2, 2P_n(x_1)(|x_2|+|x_1|)\right],
\end{eqnarray} where $a_B=\chi/me^2$ is the Bohr radius, $U[a, b, z]$ is the confluent hypergeometric function, $P_n(x_1)=\sqrt{2m\zeta_n(x_1)}$, the coefficients $C_\pm$ correspond to the domains $x_2>0$ and $x_2<0$. Even and odd solutions satisfy the boundary condition $\psi'_n(0)=0$ and $\psi_n(0)=0$, accordingly. This gives the ''molecular'' energy terms at $x_1\ll a_B$: $\zeta_0(x_1)=2\epsilon_B\log^2\frac{a_B}{|x_1|},$
\begin{eqnarray}\label{ey}
&&\zeta_n^{even}(x_1)\approx\frac{\epsilon_B}{2n^2}-\frac{\epsilon_B}{n^3\log\frac{a_B}{|x_1|}},\\
&&\zeta_n^{odd}(x_1)\approx\frac{\epsilon_B}{2n^2}-\frac{2\epsilon_B|x_1|}{n^3a_B},
\end{eqnarray} where $n\geq 1$. Using the energy terms and Eq.~(\ref{Hn}) we obtain the energy levels in the quasiclassical approximation:
\begin{eqnarray}\label{eyx}
&&\varepsilon_{00}\approx (\epsilon_B/2)\log^2\eta+...,\\
&&\varepsilon^{even}_{nN}\approx\frac{\epsilon_B}{2n^2}-
\frac{2\epsilon_B}{n^3\eta\log\frac{2\eta}{\pi^2n^3\left(N+\frac14\right)}}+...,\\
&&\varepsilon^{odd}_{nN}\approx\frac{\epsilon_B}{2n^2}-
\frac{\epsilon_B}{n^2}\left[\frac{3\pi}{4\eta^2}\left(N+\frac34\right)\right]^\frac23+...
\end{eqnarray} The distance between ''ions'' levels $\varepsilon_{n,N+1}-\varepsilon_{n,N}$ is much less, than the distance between ''electron'' levels $\varepsilon_{n+1,N}-\varepsilon_{n,N}$ by the measure of the parameter $1/\eta$.

The application of the variational approach to the Hamiltonian (\ref{Hexc2}) with total variational wave functions $\Phi_1=Be^{-\rho^2}$, $\Phi_2=Be^{-\rho}$, $\rho^2=x_1^2/a_1^2+x_2^2/a_2^2$, gives $\varepsilon_{00}=\alpha\epsilon_B\log^2\eta$, where $\alpha$ takes values $1/\pi$ and $4/\pi^2$, for $\Phi_1$ and $\Phi_2$, correspondingly. These values of $\alpha$ do not strongly differ from $1/2$ found in the exactly solvable model (see Eq.~(\ref{eyx})). The expression for $B=(2/\pi a_1a_2)^{1/2}$ holds in both models.

The binding energy of the lowest level $\varepsilon_{00}$ essentially depends on the angle $\phi_{\bf k}$, while the angular dependence of the other levels touches only ${\bf k}$-dependent corrections to these levels. The corrections have essential-singular behaviors as functions of the parameter $\eta$ different for odd and even states.

\begin{figure}
\includegraphics[width=8cm]{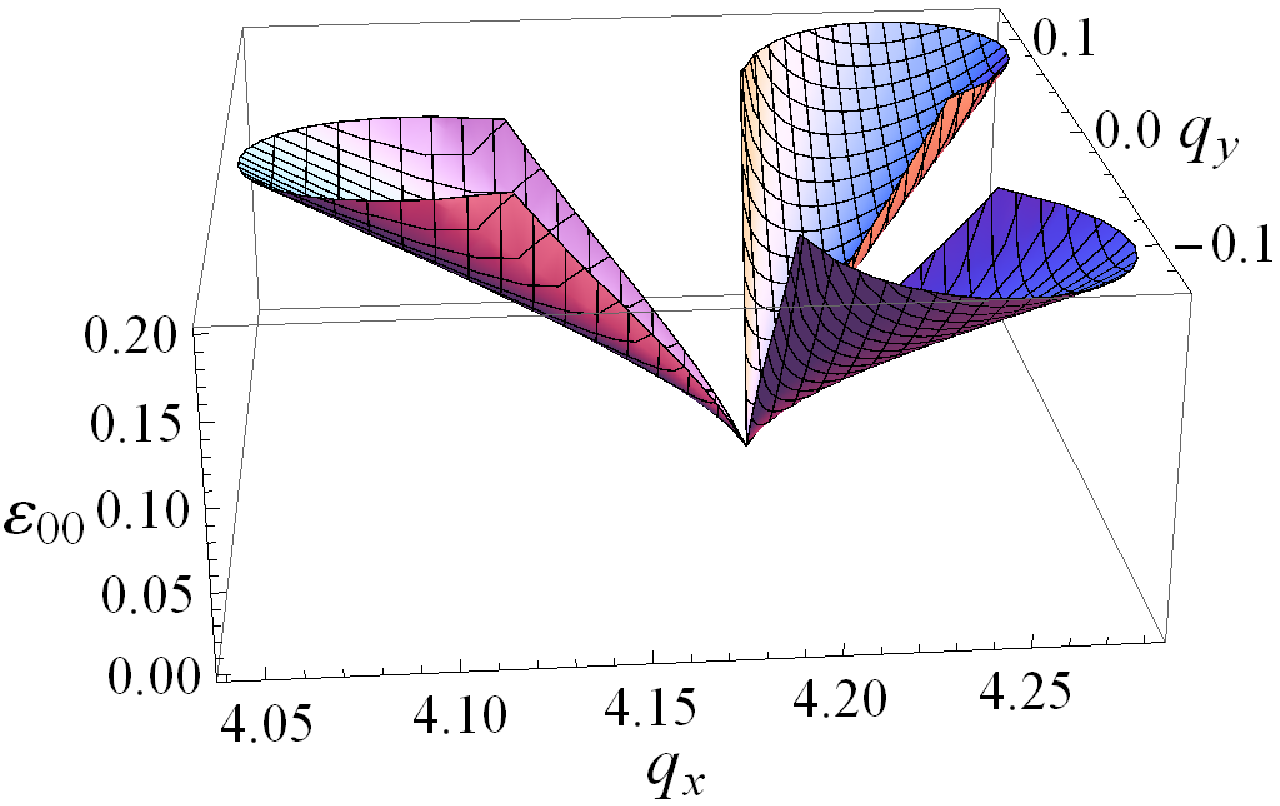}
\includegraphics[width=5.5cm]{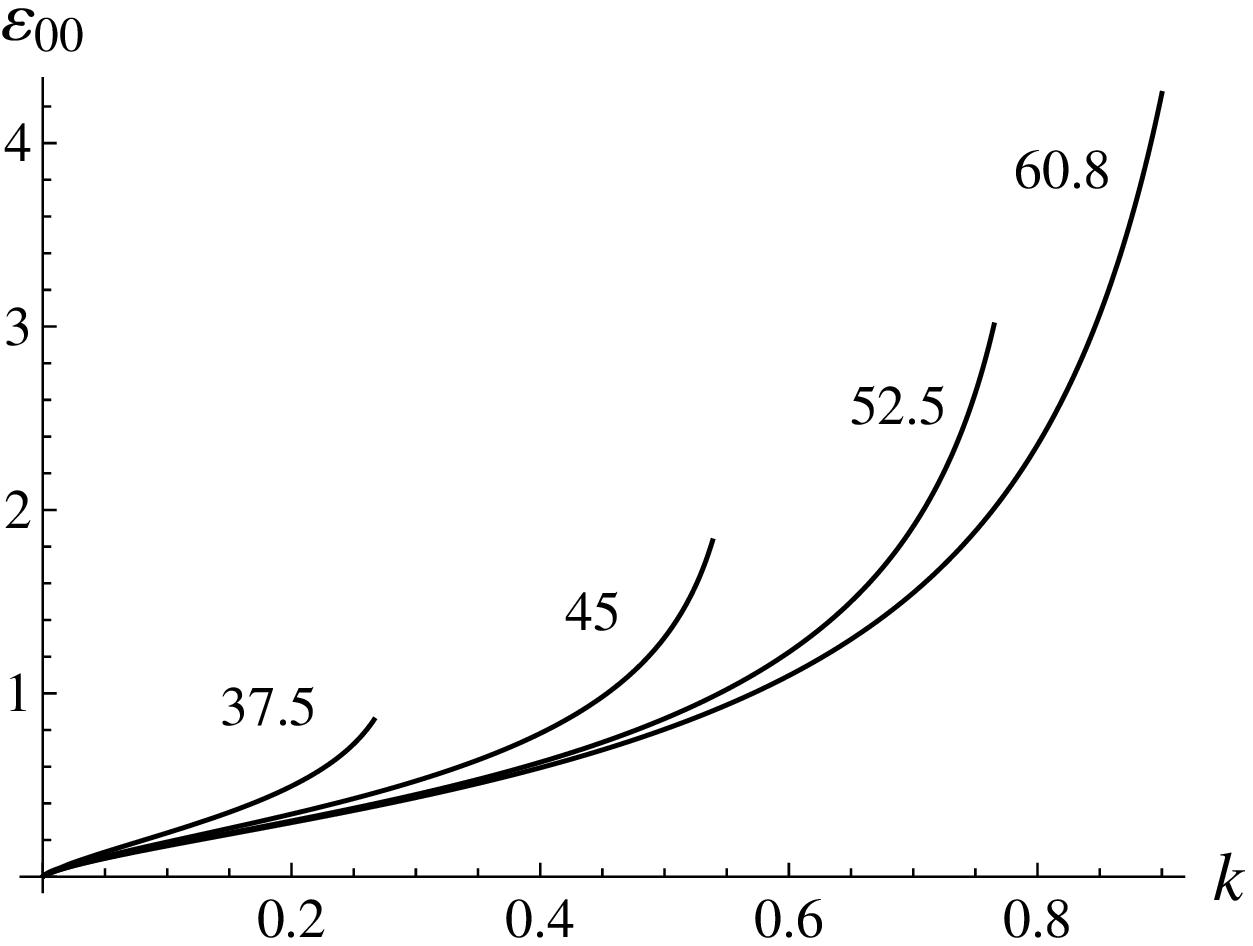}
\caption{(Color online) Top: binding energy (in eV) of the ground state of indirect exciton in graphene {\it versus} wave vector in units of reciprocal lattice constant. The exciton exists in the sectors shown in fig. \ref{fig1}. Bottom: radial sections of the top figure at fixed angles in degrees (marked). Curves run up to the ends of exciton spectrum.}
\label{fig2-3}
\end{figure}

\begin{figure}
\includegraphics[width=6.5cm]{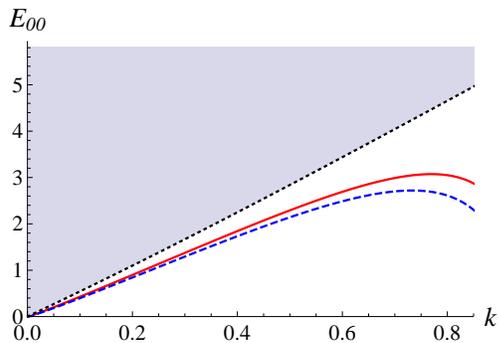}
\caption{(Color online) The total exciton energy $E_{00}$ versus $k$ along the direction 60.8 degrees. Curves are calculated using variational wave functions $\Phi_1$ (solid) and $\Phi_2$ (dashed). Filled area represents free electron and hole continuum.}
\label{fig3}
\end{figure}

We have numerically calculated the binding energy of the lowest exciton level $\varepsilon_{00}$ using the adiabatic approach. Figures~\ref{fig2-3} show the dependencies of $\varepsilon_{00}$ on the exciton momentum for $\chi=6.5$. The calculations have been done by two steps: finding the energy terms and solving the Schr\"{o}dinger equation in the $x_2$-direction. Figure~\ref{fig3} shows the total exciton energy along the medial line of the exciton existence domain.

The total exciton energy $E_{nN}$ lies in the electron-hole continuum and is always positive. However, the momentum conservation preserves the exciton from a decay, unless scattering processes (weak by assumptions) are taken into account. It should be also emphasized that the indirect exciton has two-fold degeneracy, in accordance with its total momentum. These two states have opposite symmetry in ${\bf k}$ space.

\section{Exciton absorption in graphene}

Let us discuss the possibility of optical observation of the indirect excitons in graphene. In a semiconductor with the energy gap the excitons contribute to the absorption and, especially, to the emission of light. The absence of the energy gap makes observation of the excitons in graphene a different problem from the case of a usual semiconductor, because the exciton energies are distributed between zero and several tenth of eV which smears up the exciton resonance. The large momentum of the indirect exciton blocks both the direct optical excitation and the recombination. However, the slow recombination and the inter-valley relaxation preserve the excitons (when generated someway) from the recombination or the decay. The frequency of photons with a small wave vectors, which can produce excitons, is very low. In that case, the solution can be found using a phonon-or impurity-assisted absorption or a structure where electrons obtain an additional in-plane momentum from an artificial grating or a naturally riffle of the graphene plane.

The phonon-assisted indirect exciton optical absorption has thresholds determined by one of phonon frequencies $\Omega_i({\bf K})$, where $i$ numerates the phonon type. Corresponding contribution to the absorption near the threshold $\Omega_i({\bf K})$ is determined by the exciton density of states and is proportional to $(\omega-\Omega_i+E_{nN})\theta(\omega-\Omega_i+E_{nN})$, where $\omega$ is the light frequency. This threshold behavior should be compared with more smooth threshold for free pair excitation $\propto (\omega-\Omega_i)^3\theta(\omega-\Omega_i)$. One can expect the manifestation of the exciton by an appearance of the maximum in the second derivative of absorption with respect to $\omega$ at $\omega=\Omega_i-E_{nN}$.

\section{Scattering of external electron on indirect exciton}

The other perspective way to observe excitons is based on a monolayer character of graphene. Being placed in a vacuum, the graphene layer can be studied by the scattering of vacuum electrons (forward or backward) on the graphene. The dependence of the energy loss maximum on the scattering angle gives the spectrum of the Bose excitation, in particular, excitons. The exciton spectrum gets into the operation range of the high-resolution electron energy loss spectroscopy (HREELS) with incident electron energy $\sim 10$eV \cite{hreels}. The HREELS spectra of graphene were studied in \cite{lu} (e-h pair continuum spectrum) and in \cite{koch} (plasmarons). We hope that the careful examination of the in-plane angle dependence of losses will help the identification of the exciton transitions.

The scattering of an external electron between states with the momenta ${\bf P}$ and ${\bf P}'$ on free graphene electrons is determined by the transition amplitude $A({\bf Q})$, ${\bf Q}={\bf P}-{\bf P}'$. The amplitude $A({\bf Q})$ is a product of the Coulomb factor $e^2/Q^2$ and a matrix element of $e^{i{\bf Qr}}$ between the states of the graphene electrons. For transitions between the extrema $K$ and $K'$, $A({\bf Q})=(e^2/Q^2)\int d^3r u^*_{e; K}({\bf r})u_{h; K'}({\bf r})e^{i{\bf (Q-K)r}}$, where $u_{e; K}({\bf r})$ and $u_{h; K'}({\bf r})$, are the Bloch amplitudes of the graphene electron/hole wave functions near the cone points $K$ and $K'$; note, that ${\bf K}$ is the smallest vector connecting them. The in-plane component of ${\bf Q}$ runs all vectors ${\bf K}+n{\bf a}^*+m{\bf b}^*$ shifted by the reciprocal lattice vectors. With regard to the electron-hole pairing the amplitude $A({\bf Q})$ obtains an additional multiplier, the Sommerfeld factor $F_S({\bf k})$, reflecting the value of the exciton wave function at coinciding electron and hole coordinates.

\begin{figure}
\includegraphics[width=6.5cm]{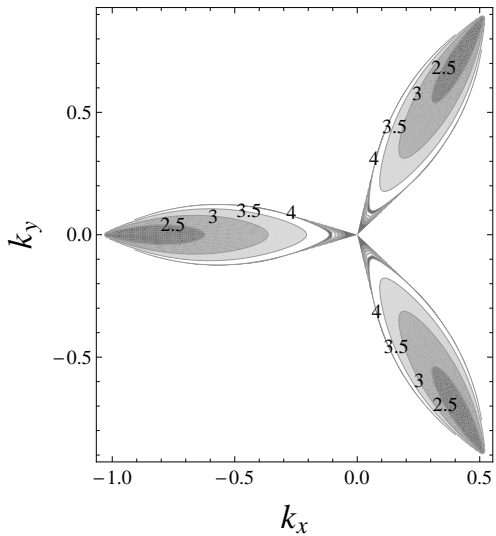}
\caption{(Color online) The map of $\ln(|F_S|^2)$ for the variational wave function $\Phi_2$.}
\label{fig4}
\end{figure}

The probability of transitions representing inelastic diffraction is determined by a sum over the reciprocal lattice vectors :
\begin{eqnarray}\label{prob}\nonumber
     W_{{\bf P}\to{\bf P}'}=2\pi|F_S({\bf k})|^2 \sum_{nm}|A({\bf Q})|^2\delta\left(\frac{P'^2-P^2}{2m_0}+E_{00}\right),
\end{eqnarray}
where $m_0$ is the bare electron mass. The energy $\delta$-function provides the main dependence of the scattering probability on the small energy and momentum transfer. Besides, the Sommerfeld factor also gives the functional dependence on the small momentum. For the variational wave functions $\Phi_1$ and $\Phi_2$, we have $F_S=B$. Figure~\ref{fig4} depicts the momentum dependence of $|F_S({\bf k})|^2$ for the variational wave function $\Phi_2$.

\begin{figure}
\includegraphics[width=3.5cm]{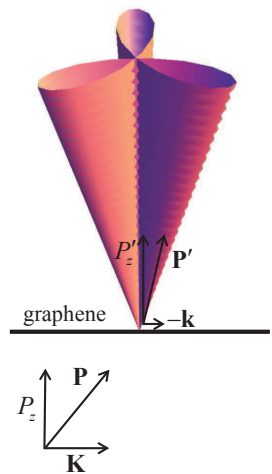}
\caption{(Color online) Sketch of proposed experiment for measurements of exciton spectrum.}
\label{fig5}
\end{figure}

Figure~\ref{fig5} shows a sketch of a proposed experiment geometry. The external electron beam with the momentum ${\bf P}=({\bf K},P_z)$, $P_z^2/2m_0\gg E_{00}$, collides with the graphene plane. Let $a^*>P>K$ (in other words, $4\pi/\sqrt{3}a>P>4\pi/3a$). In this case (together with an elastic peak ${\bf P'=P}$) the only diffraction peak $n=m=0$ is permitted and the scattered electron  momentum $({-\bf k},P_z'$), $P_z'=\sqrt{P_z^2+K^2-2m_0E_{00}-k^2}$ lies near the normal. The measurement of the normal component of the momentum $P_z'$ in the diffraction peak as a function of in-plane momentum ${-\bf k}$ gives the exciton spectrum. The effect exists if the in-plane incident momentum coincides with ${\bf K}$. That helps to identify the exciton. It should be emphasized, that the direction of the in-plane projection of ${\bf P}$ along ${\bf K}$ or $-{\bf K}$ selects one of two degenerate exciton states. As a result, the symmetry in ${\bf P}'$ space becomes trigonal.

\section{Discussion and conclusions}
The principal assumption used in the present study is the smallness of the dimensionless interaction constant $g$, which provides the smallness of the exciton binding energy as compared with its kinetic energy. The considered exciton states in graphene  are formed near the conic points.

We have based ourselves on the nearest-neighbor tight-binding model which possesses electron-hole symmetry. In fact, this symmetry is broken in graphene. The model of \cite{dress} (see, Eq. (2.21) ) accounting the asymmetry results in an additional contribution to the single-particle spectrum near the conic point: $\varepsilon_\pm\to vk-\mu k^2\cos3\phi_{\bf k}-const\cdot k^2$. However, the electron and hole contributions cancel each other in the pair energy and, hence, have no effect on the e-h binding.

The exciton states are stable towards the collisionless decay to free electron and hole pairs. At the same time, the scattering on impurities or phonons destroys the moving excitons. The weakness of this process is conditioned by the smallness of the scattering processes with participation of impurities or phonons. A rough estimate of this weakness is given by the ratio of electron scattering rate to the binding energy.

Among other kinds of scattering the Auger processes should be underlined. As the energy grows, the scattering with e-h pair production becomes prevailing in the momentum sectors where it is permitted \cite{golub}. One can see that these sectors coincides with the domains of exciton existence. The e-h scattering rate estimate gives $\nu_{e-h}\propto g^2vk$. This value should be compared with the exciton binding energy $ g^2vk \ln^2\eta$. Parametrically, the large parameter $\ln^2\eta$ results in the smallness of damping. At the same time the width of excited exciton states is comparable with their binding energy, that makes the excited exciton states unlikely observable.

Let us briefly consider the suppression of excitons in graphene due to screening by the finite concentration of equilibrium charged carriers $n_{ch}$. In the gapped semiconductors slow excitons are actual only. The equilibrium electrons and holes screen quasistatic interaction between the pair composing the exciton. Vice versa, charged carriers in graphene can follow the moving excitons only partly. The binding energy stays non-perturbed if the screening radius for the moving pair is larger than the exciton radius. The condition for exciton existence can be estimated as $n_{ch}\lsim\alpha k^2\log^2\eta$.

It should be emphasized, that in the case under consideration the total exciton energy remains positive due to the smallness of interaction and trigonal corrections. This forbids the spectrum reconstruction, such as the exciton insulator.

We have neglected the many-body corrections to single-electron spectrum. Actually, the e-e interaction leads to the electron self-energy $\tilde{g}vk\ln(1/ka)$, $\tilde{g}=g/4-g^2(5/6-\ln 2)+O(g^3)$ \cite{mishch}. Thus, we need, that not only $g\ll 1$, but typical $g\ln(1/pa)\ll 1$. The presence of this logarithmic term strongly changes the behavior of the single-electron spectrum at small $k$ and gives a negative contribution to $M$. However, this does not forbid the exciton absolutely. In fact, the interaction-induced correction to the inverse mass $1/m_{int}=-\tilde{g}v/k$ should be added to the warping correction. In a particular case of small interaction constant $g$, mass $M$ becomes negative at low $k$ and near the boundaries of the existence sectors. This narrows the range of exciton existence and shifts it from the conic point.

Another remark concerns the renormalization of the dielectric constant $\chi$, caused by the vacuum polarization of graphene \cite{kats}. The renormalization reduces the Coulomb interaction at large distance, as if the dielectric constant of external medium becomes logarithmically large. This fact improves the applicability of the weak-interaction model, even in the case of free-suspended graphene.

The excitons considered in the present paper arise near the conic points, unlike the saddle-point exciton states with energies about 5 eV studied in \cite{yang1,yang2,klit,kin,san,lee,res}. The latter, in our opinion, correspond to the e-h scattering resonance, rather than bound states. The other investigations of exciton resonances is based on the electron spectrum reconstruction caused by large interaction constant \cite{gron1,gron2}. However, the possibility of gapless excitons in a system with small $g$ considered here had not been studied earlier.

In conclusion, we have demonstrated the existence of moving indirect excitons in monolayer graphene and found their energies. The excitons are conditioned by the presence of the warping of the electron spectrum. The excitons spectrum has no gap. The many-body interaction corrections to the single-particle spectrum, in concurrence with warping, lead to the destruction of excitons with small wave vectors. However, the excitons survive if the interaction constant is small.

\section{Acknowledgments} This research was supported by the grants of RFBR No 11-02-00730 and No 11-02-12142.

\end{document}